\documentclass[preprint,nofootinbib,floats,superscriptaddress,tightenlines,amsfonts,amsmath,amssymb]{revtex4}

\usepackage{graphicx}
\usepackage{bm}
\usepackage{color}
\usepackage{multirow}
\usepackage{slashed}

\begin{document}
\baselineskip=17pt \parskip=5pt

\preprint{NCTS-PH/1905}
\hspace*{\fill}

\title{Flavor-changing hyperon decays with light invisible bosons}

\author{Gang Li}
\affiliation{Department of Physics, National Taiwan University, Taipei 106, Taiwan}

\author{Jhih-Ying Su}
\affiliation{Department of Physics, National Taiwan University, Taipei 106, Taiwan}

\author{Jusak Tandean}
\affiliation{Department of Physics, National Taiwan University, Taipei 106, Taiwan}
\affiliation{Physics Division, National Center for Theoretical Sciences, Hsinchu 300, Taiwan
\bigskip}


\begin{abstract}

We consider the strangeness-changing decays of hyperons into another baryon and missing energy
which is carried away by a pair of invisible spinless bosons.
Pursuing a model-independent approach and taking into account constraints from the kaon sector,
we find that these hyperon modes can have large rates, especially if the bosons have
parity-odd effective couplings to light quarks.
In that case the rates could attain values which are well within the expected reach of
the currently running BESIII experiment.
Since the branching fractions of hyperon decays with missing energy are tiny in the standard
model, observing them at ongoing or upcoming facilities would likely be indicative of new physics.
In addition, we show that the kaon restrictions may be weakened or absent if the bosons have
nonnegligible mass.
Specifically, if the bosons interact with the quarks solely via an axial-vector current,
in the mass region from about 115 to 175 MeV the constraints from rare kaon decays no longer
apply and a couple of the hyperon modes become the only direct probes of such interactions.
We also make comparisons with the previously explored new-physics scenario where the invisible
particles are spin-1/2 fermions.

\end{abstract}

\maketitle

\section{Introduction \label{intro}}

In the standard model (SM) the flavor-changing neutral current (FCNC) decay of a strange meson or
baryon with missing energy ($\slashed E$) in the final state proceeds mainly from the short-distance
contribution due to the quark transition \,$s\to d\nu\bar\nu$\, generated by loop diagrams \cite{Buchalla:1995vs}, with $\slashed E$ being carried away by
the unobserved neutrino pair ($\nu\bar\nu$).
Beyond the~SM there could be extra ingredients which cause changes to the SM process and/or give
rise to additional channels with one or more invisible nonstandard particles adding to~$\slashed E$.
As such decay modes of strange hadrons have very suppressed rates in the SM,
the modifications due to new physics (NP) may translate into effects big enough to be
discoverable.
These rare modes can then be expected to serve as sensitive probes of NP.

Experimentally, there has been much activity to study such processes in the meson sector.
Particularly, a while ago the E949 Collaboration \cite{Artamonov:2008qb} reported seeing
the kaon mode \,$K^+\to\pi^+\nu\bar\nu$,\, and currently there are running programs to measure it
more precisely and observe the neutral channel \,$K_L\to\pi^0\nu\bar\nu$\, by the NA62
\cite{CortinaGil:2018fkc} and KOTO \cite{Ahn:2018mvc} Collaborations, respectively.
As for other FCNC kaon decays with missing energy, the only data available are upper limits
on the branching fractions of two of the \,$K\to\pi\pi'\nu\bar\nu$\, modes from quests for them
with negative outcomes \cite{Tanabashi:2018oca,Adler:2000ic,E391a:2011aa}.
Indirectly, the corresponding limits on \,$K\to\slashed E$\, can be inferred from the existing
data on the neutral kaon visible decays \cite{Gninenko:2014sxa}.

In the baryon sector, the analogous transitions are the strangeness-changing ($|\Delta S|=1$)
hyperon decays into another baryon plus missing energy.
However, there has never been any search for them as far as we know.
Interestingly, there is a recent proposal to measure them in the BESIII
experiment~\cite{Li:2016tlt}.
Hence new data on these rare hyperon modes may be forthcoming.

It is therefore timely to investigate them theoretically to learn whether NP might impact them
significantly in light of the constraints from the kaon sector.
Initial works on the hyperon modes have been done in Refs.\,\cite{Li:2016tlt,Hu:2018luj} focusing
on NP entering through \,$s\to d\nu\bar\nu$\, interactions with the same chiral structure as in
the SM.
Subsequently, Ref.\,\cite{Tandean:2019tkm} explores a more general scenario in which
the underlying NP operators involve other Lorentz structures and the pair of invisible particles
could be nonstandard spin-1/2 fermions.
In this more general possibility, it turns out that the hyperon decays could have rates which
are greatly increased relative to the SM expectations, even up to levels potentially
detectible by BESIII in the near future \cite{Tandean:2019tkm}.

In the present paper, we extend the analysis of Ref.\,\cite{Tandean:2019tkm} and entertain
the possibility that the invisible pair consists of a spin-0 boson, $\phi$, and its antiparticle,
$\bar\phi$, rather than fermions.
We assume that the $\phi$ field is complex and does not carry any SM gauge charge but is either
charged under some symmetry of a dark sector beyond the SM or odd under a $Z_2$ symmetry
which does not influence SM fields.
Moreover, we suppose that any other states beyond the SM are heavier than the weak scale.
As a consequence, $\phi$ and $\bar\phi$ always appear together in their interactions with SM quarks
at low energies.
Scenarios where such nonstandard bosons have $ds$ couplings that contribute to kaon processes
have been addressed before in the literature under specific NP contexts \cite{Bird:2004ts,
Bird:2006jd,He:2010nt,Gninenko:2015mea}.
Here we pursue a model-independent approach, starting from the most general effective Lagrangian
that satisfies SM gauge invariance and contains operators describing $ds\phi\phi$
interactions at leading order.

We will show that the rates of \,$|\Delta S|=1$\, hyperon decays with missing energy can be
sizable in the presence of the new interactions, similarly to what happens in the invisible
fermion case studied in Ref.\,\cite{Tandean:2019tkm}.
Such a possibility is partly attributable to the fact that these hyperon transitions and their
kaon counterparts do not probe the same set of the underlying NP operators.
In particular, \,$K\to\pi\slashed E$\, is sensitive exclusively to parity-even $ds\phi\phi$
operators and \,$K\to\slashed E$\, to the parity-odd ones, while \,$K\to\pi\pi'\slashed E$\,
and the spin-1/2 hyperon modes can probe both.
Another factor is the kaon data situation at the moment, which has a bearing on limiting mainly
the parity-even operators, due to the existing empirical information
on \,$K\to\pi\nu\bar\nu$\, decays \cite{CortinaGil:2018fkc,Ahn:2018mvc,Tanabashi:2018oca}.
By contrast, the restraints from \,$K\to\slashed E$ and \,$K\to\pi\pi'\slashed E$\, are
relatively far weaker and still permit the parity-odd operators to bring about considerable
amplifying effects on the hyperon rates.

Besides the similarities, there are differences between the bosonic and fermionic cases.
Due to the invisible particle having spin 1/2, the latter scenario involves more operators
with their associated coupling parameters, leading to more possibilities in the phenomenology,
especially if the invisible particle's mass is not negligible.
Our examples will illustrate the salient differences and similarities.

The structure of the remainder of the paper is as follows.
In the next section we write down the $ds\phi\phi$ operators of interest,
treating them in a model-independent manner.
In Sec.\,\ref{amps} we first deal with the FCNC hyperon decays with missing energy,
deriving their differential rates, and subsequently provide the formulas for
the corresponding kaon decays of concern.
In Sec.\,\ref{numeric} we present our numerical results.
Evaluating the hyperon modes and taking into account restrictions from the kaon sector, we
demonstrate that the hyperon rates can be substantial and within the proposed BESIII reach.
We also make comparisons with the fermionic case examined in Ref.\,\cite{Tandean:2019tkm}, where
in the numerical analysis the invisible fermion's mass was assumed to be negligible or zero for
simplicity, and furthermore discuss what may happen if it is not negligible.
In the latter case, we point out that under certain conditions a few of the hyperon modes are
allowed to have rates higher than their counterparts computed in Ref.\,\cite{Tandean:2019tkm}.
In Sec.\,\ref{concl} we give our conclusions.

\section{Interactions\label{Lf}}

At lowest order, the effective operators responsible for the SM-gauge-invariant interactions of
$\phi$ with quarks as specified above are of mass dimension six~\cite{Badin:2010uh,Kamenik:2011vy}.
The relevant Lagrangian ${\cal L}_{\textsc{np}}$ is given by \cite{Kamenik:2011vy}
\begin{align} \label{Lnp0}
-\Lambda_{\textsc{np}}^2\, {\cal L}_{\textsc{np}}^{} & \,=\,
\Big( {\cal C}_{jk}^{}\, \overline{q_j^{}}\gamma^\eta q_k^{}
+ {\cal C}_{jk}'\, \overline{d_j}\gamma^\eta d_k^{} \Big) i
\big(\phi^\dagger\partial_\eta\phi-\partial_\eta\phi^\dagger\phi\big)
+ \Big( {\cal C}_{jk}''\, \overline{q_j^{}} d_k^{} H
+ {\rm H.c.} \Big) \phi^\dagger \phi \,, &
\end{align}
where $\Lambda_{\textsc{np}}$ is a heavy mass scale representing the underlying NP
interactions, ${\cal C}_{jk}^{(\prime,\prime\prime)}$ in our model-independent approach
are dimensionless free parameters which are generally complex, $q_j^{}$ and $d_j^{}$ denote
a left-handed quark doublet and right-handed down-type quark singlet, respectively, $H$~stands for
the Higgs doublet, and summation over family indices \,$j,k=1,2,3$\, is implicit.\footnote{If
the $\phi$ field is real instead, the ${\cal C}_{jk}^{}$ and ${\cal C}_{jk}'$ terms are absent
from Eq.\,(\ref{Lnp0}).\medskip}
The Hermiticity of ${\cal L}_{\textsc{np}}$ implies
\,${\cal C}_{jk}^{(\prime)}={\cal C}_{kj}^{(\prime)*}$.\,

To deal with the contributions of ${\cal L}_{\textsc{np}}$ to the hyperon and kaon decays, it is
convenient to rewrite the $ds\phi\phi$ operators in it explicitly separating their parity-even
and -odd parts.
Thus, choosing to work in the mass basis of the down-type quarks, at low energies we can express
\begin{align} \label{Lnp}
-{\cal L}_{\textsc{np}}^{} & \,\supset\, \big(\textsf c_\phi^{\texttt V}\,\overline{d}\gamma^\eta s
+ \textsf c_\phi^{\texttt A}\, \overline{d}\gamma^\eta\gamma_5^{}s \big)
i\big(\phi^\dagger\partial_\eta\phi-\partial_\eta\phi^\dagger\phi\big)
+ \big( \textsf c_\phi^{\texttt S}\, \overline{d}s + \textsf c_\phi^{\texttt P}\,
\overline{d}\gamma_5^{}s \big) \phi^\dagger \phi \;+\; {\rm H.c.} \,, &
\end{align}
where
\begin{align}
\textsf c_\phi^{\texttt V} & \,=\,
\frac{{\cal C}_{12}^{}+{\cal C}_{12}'}{2\Lambda_{\textsc{np}}^2} \,, &
\textsf c_\phi^{\texttt A} & \,=\,
\frac{-{\cal C}_{12}^{}+{\cal C}_{12}'}{2\Lambda_{\textsc{np}}^2} \,, &
\textsf c_\phi^{\texttt S} & \,=\,
\frac{{\cal C}_{12}''+{\cal C}_{21}^{\prime\prime*}}{2\sqrt2\,\Lambda_{\textsc{np}}^2}\, v \,, &
\textsf c_\phi^{\texttt P} & \,=\,
\frac{{\cal C}_{12}''-{\cal C}_{21}^{\prime\prime*}}{2\sqrt2\,\Lambda_{\textsc{np}}^2}\, v \,,
\end{align}
with \,$v\simeq246$\,GeV\, being the Higgs vacuum expectation value.\footnote{We mention that for
effective flavor-changing operators involving light quarks and SM-gauge-singlet dark particles
the running of their coefficients from high to low energies has been estimated to be
negligible~\cite{Arteaga:2018cmw}.}
As we will see below, in the decay rates the contributions of
$\textsf c_\phi^{\texttt V,\texttt A,\texttt S,\texttt P}$ do not interfere.
In numerical calculations, we will adopt the phenomenological viewpoint that these free parameters
can have any values compatible with the empirical constraints and perturbativity,
implying that some of them may be taken to be vanishing or much less than the others.

\section{Amplitudes and rates\label{amps}}

\subsection{Baryon decays\label{hyperons}}

Our hyperon decays of interest are
\,$\mathfrak B\to\mathfrak B'\phi\bar\phi$\, with
\,$\mathfrak{BB}'=\Lambda n,\Sigma^+p,\Xi^0\Lambda,\Xi^0\Sigma^0,\Xi^-\Sigma^-$,\, all of
the baryons having spin 1/2, and \,$\Omega^-\to\Xi^-\phi\bar\phi$\, where $\Omega^-$ has spin 3/2.
One or more of these modes may be searched for at BESIII which can produce copious
$\Lambda$, $\Sigma$, $\Xi$, and $\Omega$ hyperons \cite{Li:2016tlt}.

To derive the amplitudes for these decays, we need the baryonic matrix elements of the quark
portions of the operators in Eq.\,(\ref{Lnp}).
They can be estimated with the aid of flavor-SU(3) chiral perturbation theory at
leading order \cite{He:2005we,Tandean:2019tkm}.
For \,$\mathfrak B\to\mathfrak B'\phi\bar\phi$,\, from Ref.\,\cite{Tandean:2019tkm} we have
\begin{align} \label{<B'B>}
\big\langle\mathfrak B'\big|\overline{d}\gamma^\eta s\big|\mathfrak B\big\rangle & \,=\,
{\cal V}_{\mathfrak B'\mathfrak B}^{}\,\bar u_{\mathfrak B'}^{}\gamma^\eta u_{\mathfrak B}^{} \,,
~ & \big\langle{\mathfrak B'}\big|\overline{d}\gamma^\eta\gamma_5^{}s\big|{\mathfrak B}\big\rangle
& \,=\, \bar u_{\mathfrak B'}^{} \bigg( \gamma^\eta {\cal A}_{\mathfrak B'\mathfrak B}^{}
- \frac{{\cal P}_{\mathfrak B'\mathfrak B}}{B_0}\, \texttt Q^\eta{} \bigg)
\gamma_5^{} u_{\mathfrak B}^{} \,,
\nonumber \\
\big\langle{\mathfrak B}'\big|\overline{d}s\big|{\mathfrak B}\big\rangle & \,=\,
{\cal S}_{\mathfrak B'\mathfrak B}^{}\, \bar u_{\mathfrak B'}^{} u_{\mathfrak B}^{} \,, &
\big\langle{\mathfrak B'}\big|\overline{d} \gamma_5^{}s\big|{\mathfrak B}\big\rangle
& \,=\, {\cal P}_{\mathfrak B'\mathfrak B}^{}\,
\bar u_{\mathfrak B'}^{}\gamma_5^{}u_{\mathfrak B}^{} \,,
\end{align}
where
\,${\cal V}_{\mathfrak B'\mathfrak B}=-3/\sqrt{6},-1,3/\sqrt{6},-1/\sqrt2,1$\, and
\,${\cal A}_{\mathfrak B'\mathfrak B}=-(D+3F)/\sqrt6,D-F,(3F-D)/\sqrt6$, $-(D+F)/\sqrt2,D+F$\,
for \,$\mathfrak{BB}'=\Lambda n,\Sigma^+p,\Xi^0\Lambda,\Xi^0\Sigma^0,\Xi^-\Sigma^-$,\,
respectively, $\bar u_{\mathfrak B'}$ and $u_{\mathfrak B}$ denote the Dirac spinors of
the baryons, \,$\texttt Q=p_{\mathfrak B}^{}-p_{\mathfrak B'}^{}$,\, with $p_{\mathfrak B}^{}$
and $p_{\mathfrak B'}^{}$ representing their four-momenta,
\,${\cal S}_{\mathfrak B'\mathfrak B}={\cal V}_{\mathfrak B'\mathfrak B}^{}
(m_{\mathfrak B}^{}-m_{\mathfrak B'}^{})/(m_s^{}-\hat m)$,\,
and
\,${\cal P}_{\mathfrak B'\mathfrak B}^{}={\cal A}_{\mathfrak B'\mathfrak B\,}^{} B_0^{}
(m_{\mathfrak B'}^{}+m_{\mathfrak B}^{})/\big(m_K^2-\texttt Q^2\big)$,\,
with $m_{\mathfrak B{}^{(\prime)}}$ and $m_K$ in ${\cal S}_{\mathfrak B'\mathfrak B}$
and ${\cal P}_{\mathfrak B'\mathfrak B}$ referring to the isospin-averaged masses of
the baryon $\mathfrak B{}^{(\prime)}$ and kaon, \,$\hat m=(m_u^{}+m_d^{})/2$\,
and $m_s^{}$ being light quarks' masses, and \,$B_0^{}=m_K^2/(\hat m+m_s^{})$.\,
For \,$\Omega^-\to\Xi^-\phi\bar\phi$,\, the hadronic matrix elements that do not vanish at
lowest order in chiral perturbation theory are~\cite{Tandean:2019tkm}
\begin{align} \label{<XO>}
\langle\Xi^-\big|\overline{d}\gamma^\eta\gamma_5^{}s|\Omega^-\rangle & \,=\, {\cal C}\,
\bar u_{\Xi}^{} \Bigg( u_{\Omega}^\eta + \frac{\tilde{\textsc q}{}^\eta\,
\tilde{\textsc q}{}_\kappa^{}}{m_K^2-\tilde{\textsc q}{}^2}\, u_\Omega^\kappa \Bigg) , &
\langle\Xi^-|\overline{d}\gamma_5^{}s|\Omega^-\rangle & \,=\, \frac{B_0^{}\,{\cal C}\,
\tilde{\textsc q}_\kappa^{}}{\tilde{\textsc q}{}^2-m_K^2}\,\bar u_{\Xi}^{}u_\Omega^\kappa \,,
\end{align}
where \,$\tilde{\textsc q}=p_{\Omega^-}^{}-p_{\Xi^-}^{}$\, and
$u_\Omega^\eta$ is a Rarita-Schwinger spinor.
The constants $D$, $F$, and $\cal C$ above come from the lowest-order
chiral Lagrangian and can be fixed from baryon decay data.
In numerical analysis, we will employ the same (central) values of these and other input parameters
as those given in Ref.\,\cite{Tandean:2019tkm}.

With Eqs.\,\,(\ref{<B'B>}) and (\ref{<XO>}), we derive the amplitudes for
\,$\mathfrak B\to\mathfrak B'\phi\bar\phi$\, and \,$\Omega^-\to\Xi^-\phi\bar\phi$\, to be
\begin{align}
{\cal M}_{\mathfrak B\to\mathfrak B'\phi\bar\phi}^{} & \,=\,
\bar u_{\mathfrak B'}^{} \Big[ \gamma^\eta(\textsl{\texttt p}-\bar{\textsl{\texttt p}})_\eta^{}
\Big( \textsf c_\phi^{\texttt V} {\cal V}_{\mathfrak B'\mathfrak B}^{}
+ \gamma_5^{}\, \textsf c_\phi^{\texttt A\;} {\cal A}_{\mathfrak B'\mathfrak B}^{} \Big)
+ \textsf c_\phi^{\texttt S}\, {\cal S}_{\mathfrak B'\mathfrak B}^{}
+ \gamma_5^{}\, \textsf c_\phi^{\texttt P}\, {\cal P}_{\mathfrak B'\mathfrak B}^{}
\Big] u_{\mathfrak B}^{} \,,
\nonumber \\
{\cal M}_{\Omega\to\Xi\phi\bar\phi}^{} & \,=\, \Bigg[ \textsf c_\phi^{\texttt A}\,
(\textsl{\texttt p}-\bar{\textsl{\texttt p}})_\eta^{} + \frac{B_0^{}\, \textsf c_\phi^{\texttt P}\,
(\textsl{\texttt p}+\bar{\textsl{\texttt p}})_\eta^{}} {(\textsl{\texttt p}
+ \bar{\textsl{\texttt p}})^2-m_K^2} \Bigg] {\cal C}\, \bar u_\Xi^{} u_\Omega^\eta \,,
\end{align}
where $\textsl{\texttt p}$ and $\bar{\textsl{\texttt p}}$ denote the momenta of $\phi$ and
$\bar\phi$, respectively.
Subsequently, we arrive at the differential decay rates
\begin{align} \label{Gamma'}
\frac{d\Gamma_{{\mathfrak B}\to{\mathfrak B}'\phi\bar\phi}^{}}{d\hat s} & \,=\, \frac{\beta\,
\lambda_{\mathfrak B\mathfrak B'}^{1/2}}{256 \pi^3 m_{\mathfrak B}^3} \begin{array}[t]{l} \! \bigg[
\displaystyle \beta^2 \bigg( \frac{\lambda_{\mathfrak B\mathfrak B'}}{3}
+ \tilde\sigma_{\mathfrak B\mathfrak B'}^- \hat s \bigg)
\big|\textsf c_\phi^{\texttt V}\big|\raisebox{2pt}{$^2$} {\cal V}_{\mathfrak B'\mathfrak B}^2 + \beta^2
\bigg( \frac{\lambda_{\mathfrak B\mathfrak B'}}{3} + \tilde\sigma_{\mathfrak B\mathfrak B'}^+\hat s
\bigg) \big|\textsf c_\phi^{\texttt A}\big|\raisebox{2pt}{$^2$} {\cal A}_{\mathfrak B'\mathfrak B}^2
\vspace{3pt} \\ +\;
\tilde\sigma_{\mathfrak B\mathfrak B'}^+\, \big|\textsf c_\phi^{\texttt S}\big|\raisebox{2pt}{$^2$}
{\cal S}_{\mathfrak B'\mathfrak B}^2 + \tilde\sigma_{\mathfrak B\mathfrak B'}^-\,
\big|\textsf c_\phi^{\texttt P}\big|\raisebox{2pt}{$^2$}\, {\cal P}_{\mathfrak B'\mathfrak B}^2
\bigg] \,, \end{array}
\nonumber \\
\frac{d\Gamma_{\Omega\to\Xi\phi\bar\phi}^{}}{d\hat s} & \,=\, \frac{\beta\,
\lambda_{\Omega^-\Xi^-}^{1/2}\,\tilde\sigma_{\Omega^-\Xi^-\,}^+{\cal C}^2}{3072\pi^3m_{\Omega^-}^5}
\Bigg[ \beta^2 \bigg(\frac{\lambda_{\Omega^-\Xi^-}}{3}+4m_{\Omega^-}^2\hat s\bigg)
\big|\textsf c_\phi^{\texttt A}\big|\raisebox{2pt}{$^2$} + \frac{\lambda_{\Omega^-\Xi^-\,}^{}B_0^2}
{\big(m_K^2-\hat s\big)\raisebox{2pt}{$^2$}}
\big|\textsf c_\phi^{\texttt P}\big|\raisebox{2pt}{$^2$} \Bigg] \,,
\end{align}
where
\begin{align}
\beta & \,=\, \sqrt{1-\frac{4 m_\phi^2}{\hat s}} \,, &
\lambda_{XY}^{} & \,=\, {\cal K}\big(m_X^2,m_Y^2,\hat s\big) \,, &
{\cal K}(x,y,z) & \,=\, (x-y-z)^2-4y z \,,
\nonumber \\
\hat s & \,=\, (\textsl{\texttt p}+\bar{\textsl{\texttt p}})^2 \,, &
\tilde\sigma_{XY}^\pm & \,=\, (m_X^{}\pm m_Y^{})^2 - \hat s \,.
\end{align}
To get the decay rates, we integrate the differential rates in Eq.\,(\ref{Gamma'})
over \,$4m_\phi^2\le\hat s\le(m_{\mathfrak B}-m_{\mathfrak B'})^2$\, and
\,$4m_\phi^2\le\hat s\le(m_{\Omega^-}-m_{\Xi^-})^2$,\, respectively.

Following Ref.\,\cite{Tandean:2019tkm}, in our numerical treatment of these hyperon decay rates
we will incorporate form-factor effects not yet taken into account in Eqs.\,\,(\ref{<B'B>}) and
(\ref{<XO>}).
Specifically, we will make the changes
\,${\cal V}_{\mathfrak B'\mathfrak B} \to
\big(1+2\texttt Q^2/M_V^2\big){\cal V}_{\mathfrak B'\mathfrak B}$\,
and
\,${\cal A}_{\mathfrak B'\mathfrak B} \to
\big(1+2\texttt Q^2/M_A^2\big){\cal A}_{\mathfrak B'\mathfrak B}$\,
with \,$M_V=0.97(4)$\,GeV and \,$M_A=1.25(15)$\,GeV,\, which are in line with the commonly
used parametrization in experimental studies of hyperon semileptonic
decays \cite{Bourquin:1981ba,Gaillard:1984ny,Hsueh:1988ar,Dworkin:1990dd,Batley:2006fc}.
Moreover, for the $\Omega^-$ decay we apply the modification
\,${\cal C}\to{\cal C}/\big(1-\tilde{\textsc q}{}^2/M_A^2\big)\raisebox{1pt}{$^2$}$.\,

\subsection{Kaon decays\label{kaons}}

For \,$K\to\phi\bar\phi$\, and \,$K\to\pi\phi\bar\phi$,\, the mesonic matrix elements
that do not vanish are given by
\begin{align}
\langle0|\overline{d}\gamma^\eta\gamma_5^{}s|\overline K{}^0\rangle & =
\langle0|\overline{s}\gamma^\eta\gamma_5^{}d|K^0\rangle =\, -i f_K^{} p_K^\eta \,, ~~~ ~~~~
\langle0|\overline{d}\gamma_5^{}s|\overline K{}^0\rangle =
\langle0|\overline{s}\gamma_5^{}d|K^0\rangle =\, i B_0^{}f_K^{} \,,
\nonumber \\
\langle\pi^-|\bar d\gamma^\eta s|K^-\rangle & \,=\,
\big(p_K^\eta+p_\pi^\eta\big) f_+^{} \,+\, \big(f_0^{}-f_+^{}\big) q_{K\pi}^\eta\,
\frac{m_K^2-m_\pi^2}{q_{K\pi}^2} \,,
\nonumber \\ \label{<piK>}
\langle\pi^-|\bar d s|K^-\rangle & \,=\, B_0^{} f_0^{} \,,
~~~ ~~~~ ~~~ q_{K\pi}^{} \,=\, p_K^{}-p_\pi^{} \,,
\end{align}
where \,$f_K^{}=155.6(4)$\,MeV \cite{Tanabashi:2018oca} is the kaon decay constant and
$f_{+,0}^{}$ are form factors dependent on $q_{K\pi}^2$.
We also have
\,$\big\langle\pi^0\big|\bar d(\gamma^\eta,1)s\big|\overline K{}^0\big\rangle =
\big\langle\pi^0\big|\bar s(-\gamma^\eta,1)d\big|K^0\big\rangle =
-\big\langle\pi^-\big|\bar d(\gamma^\eta,1)s\big|K^-\big\rangle/\sqrt2$\,
and
\,$\big\langle\pi^-\big|\bar d\gamma^\eta s\big|K^-\big\rangle=
\big\langle\pi^+\big|\bar u\gamma^\eta s\big|\overline K{}^0\big\rangle$\,
under the assumption of isospin symmetry.
We can then employ
\,$f_{+,0}^{}=\textsf f_+^{}(0)\big(1+\lambda_{+,0}^{}\,q_{K\pi}^2/m_{\pi^+}^2\big)$\,
with \,$\lambda_+^{}=0.0271(10)$\, and \,$\lambda_0^{}=0.0142(23)$\, from
\,$K_L\to\pi^+\mu^-\nu$ data \cite{Tanabashi:2018oca} and \,$\textsf f_+^{}(0)=0.9681(23)$\,
from lattice work \cite{Charles:2015gya}.
For \,$K^-\to\pi^0\pi^-\phi\bar\phi$\, and \,$K_L\to\pi^0\pi^0\phi\bar\phi$\, the pertinent
matrix elements are \cite{Tandean:2019tkm}
\begin{align} \label{<K->pp>}
\big\langle\pi^0(p_0^{})\,\pi^-(p_-^{})\big|\bar d\big(\gamma^\eta,1\big)\gamma_5^{}s
\big|K^-\big\rangle
& \,=\, \frac{i\sqrt2}{f_K^{}} \bigg[ \big( p_0^\eta - p_-^\eta, 0 \big)
+ \frac{(p_0^{}-p_-^{})\cdot\tilde{\textsl{\texttt q}}}{m_K^2-\tilde{\textsl{\texttt q}}{}^2}
\bigl( \tilde{\textsl{\texttt q}}{}^\eta, -B_0^{} \bigr) \bigg] \,, &
\nonumber \\
\big\langle\pi^0(p_1^{})\,\pi^0(p_2^{})\big|\bar d\big(\gamma^\eta,1\big)\gamma_5^{}s
\big|\,\overline{\!K}{}^0\big\rangle
& \,=\, \frac{i}{f_K^{}} \bigg[ \big( p_1^\eta + p_2^\eta, 0 \big)
+ \frac{(p_1^{}+p_2^{})\cdot\tilde{\textsl{\texttt q}}}{m_K^2-\tilde{\textsl{\texttt q}}{}^2}
\bigl( \tilde{\textsl{\texttt q}}{}^\eta, -B_0^{} \bigr) \bigg] \,, &
\end{align}
where \,$\tilde{\textsl{\texttt q}}=p_{K^-}^{}-p_0^{}-p_-^{}=p_{\bar K^0}^{}-p_1^{}-p_2^{}$.\,
In the $K^-$ case, we will drop the small contribution
\,$\langle\pi^0\pi^-|\bar d\gamma^\eta s|K^-\rangle\neq0$\, arising from the anomaly
Lagrangian \cite{Kamenik:2011vy} which is at next-to-leading order in the chiral expansion.
We also ignore form-factor effects in estimating the \,$K\to\pi\pi'\phi\bar\phi$\, rates.

From Eq.\,(\ref{<piK>}), we obtain the amplitudes for \,$K_{L,S}\to\phi\bar\phi$\, induced by
${\cal L}_\phi$ to be
\begin{align} \label{MK2ff}
{\cal M}_{K_L^{}\to\phi\bar\phi}^{} & \,=\,
-\sqrt2\, B_0^{}\,f_K^{}\, {\rm Im}\,\textsf c_\phi^{\texttt P} \,, &
{\cal M}_{K_S^{}\to\phi\bar\phi}^{} & \,=\,
i\sqrt2\, B_0^{}\, f_K^{}\, {\rm Re}\,\textsf c_\phi^{\texttt P} \,, ~~~
\end{align}
with the approximate relations \,$\sqrt2\, K_{L,S}^{}=K^0\pm\overline K{}^0$.\,
The corresponding decay rates are
\begin{align} \label{GK2ff}
\Gamma_{K_L^{}\to\phi\bar\phi}^{} & \,=\, \frac{B_0^2 f_K^2}{8\pi m_{K^0}^{}}
\sqrt{1-\frac{4m_\phi^2}{m_{K^0}^2}}\,
\big({\rm Im}\,\textsf c_\phi^{\texttt P}\big)\raisebox{2pt}{$^2$} \,, &
\Gamma_{K_S^{}\to\phi\bar\phi}^{} & \,=\,
\frac{\big({\rm Re}\,\textsf c_\phi^{\texttt P}\big)\raisebox{1pt}{$^2$}}
{\big({\rm Im}\,\textsf c_\phi^{\texttt P}\big)\raisebox{2pt}{$^2$}}\,
\Gamma_{K_L^{}\to\phi\bar\phi}^{} \,. ~~~
\end{align}
For the three-body modes, we derive
\begin{align} \label{K2pff}
{\cal M}_{K^-\to\pi^-\phi\bar\phi}^{} & \,=\, 2 f_+^{}\,
p_K^{}\!\cdot\!(\textsl{\texttt p}-\bar{\textsl{\texttt p}})\, \textsf c_\phi^{\texttt V}
+ B_0^{} f_0^{}\, \textsf c_\phi^{\texttt S} \,,
\\
{\cal M}_{K_L^{}\to\pi^0\phi\bar\phi}^{} & \,=\, -2i f_+^{}\, p_K^{}\!\cdot\!
(\textsl{\texttt p}-\bar{\textsl{\texttt p}})\, {\rm Im}\, \textsf c_\phi^{\texttt V}
\,-\, B_0^{} f_0^{}\, {\rm Re}\, \textsf c_\phi^{\texttt S} \,. &
\end{align}
These lead to the differential decay rates
\begin{align}
\frac{d\Gamma_{K^-\to\pi^-\phi\bar\phi}^{}}{d\hat s} & \,=\, \frac{\beta\, \lambda_{K^-\pi^-}^{1/2}}
{768 \pi^3 m_{K^-}^3} \Big( \beta^2 \lambda_{K^-\pi^-\,}^{} f_+^2\,
\big|\textsf c_\phi^{\texttt V}\big|\raisebox{1pt}{$^2$} + 3 B_0^2 f_0^2\,
\big|\textsf c_\phi^{\texttt S}\big|\raisebox{1pt}{$^2$} \Big) \,,
\\
\frac{d\Gamma_{K_L^{}\to\pi^0\phi\bar\phi}^{}}{d\hat s} & \,=\, \frac{\beta\,
\lambda_{K^0\pi^0}^{1/2}}{768 \pi^3 m_{K^0}^3} \Big[ \beta^2 \lambda_{K^0\pi^0\,}^{} f_+^2\,
\big({\rm Im}\,\textsf c_\phi^{\texttt V}\big)\raisebox{1pt}{$^2$} + 3 B_0^2 f_0^2\,
\big({\rm Re}\,\textsf c_\phi^{\texttt S}\big)\raisebox{1pt}{$^2$} \Big] \,, &
\end{align}
and the rates result from integrating them over
\,$4m_\phi^2\le\hat s\le(m_{K^-,K^0}-m_{\pi^-,\pi^0})^2$,\,  respectively.
For the four-body kaon decays, in view of Eq.\,(\ref{<K->pp>}), the amplitudes are
\begin{align}
{\cal M}_{K^-\to\pi^0\pi^-\phi\bar\phi}^{} & \,=\, \frac{i\sqrt2}{f_K^{}} \Bigg[
\textsf c_\phi^{\texttt A}\, \big(p_0^{}-p_-^{}\big)\!\cdot\!
(\textsl{\texttt p}-\bar{\textsl{\texttt p}}) - B_0^{}\, \textsf c_\phi^{\texttt P}\,
\frac{(p_0^{}-p_-^{})\cdot\tilde{\textsl{\texttt q}}}{m_K^2-\hat s} \Bigg] \,,
\\
{\cal M}_{K_L^{}\to\pi^0\pi^0\phi\bar\phi}^{} & \,=\,
\frac{i\sqrt2}{f_K^{}} \Bigg[ {\rm Re}\,\textsf c_\phi^{\texttt A}\, (p_1^{}+p_2^{})\!\cdot\!
(\textsl{\texttt p}-\bar{\textsl{\texttt p}}) - i B_0^{}\, {\rm Im}\,\textsf c_\phi^{\texttt P}\,
\frac{(p_1^{}+p_2^{})\cdot\tilde{\textsl{\texttt q}}}{m_K^2-\hat s} \Bigg] \,, &
\end{align}
from which we arrive at the double-differential decay rates\footnote{These formulas in
Eq.\,(\ref{ddr}) agree with those given in Ref.\,\cite{Kamenik:2011vy}, except for the overall
factor \,$\beta=\big(1-4 m_\phi^2/\hat s\big){}^{1/2}$ which appears to be missing from the latter.}
\begin{align} \label{ddr}
\frac{d^2\Gamma_{K^-\to\pi^0\pi^-\phi\bar\phi}^{}}{d\hat s\,d\hat\varsigma} & \,=\,
\frac{\beta\, \beta_{\hat\varsigma}^3\, \tilde{\lambda}_{K^-}^{1/2}}{72(4\pi)^5 f_K^2 m_{K^-}^3}
\Bigg[ \beta^2\, \big(\tilde\lambda_{K^-} + 12 \hat s \hat\varsigma\big)\,
\big|\textsf c_\phi^{\texttt A}\big|^2 + \frac{3 \tilde\lambda_{K^-}\, B_0^2}
{\big(m_K^2-\hat s\big)\raisebox{1pt}{$^2$}} \big|\textsf c_\phi^{\texttt P}\big|^2 \Bigg] \,,
\nonumber \\
\frac{d^2\Gamma_{K_L^{}\to\pi^0\pi^0\phi\bar\phi}^{}}{d\hat s\,d\hat\varsigma} & \,=\,
\frac{\beta\,\beta_{\hat\varsigma}^{}\,\tilde{\lambda}_{K^0}^{1/2}}{48(4\pi)^5f_K^2m_{K^0}^3}\Bigg[
\beta^2\,\tilde\lambda_{K^0}^{}\, \big({\rm Re}\,\textsf c_\phi^{\texttt A}\big)\raisebox{1pt}{$^2$}
+ \frac{3 B_0^2\, \big(m_{K^0}^2-\hat s-\hat\varsigma\big)\raisebox{1pt}{$^2$}}
{\big(m_K^2-\hat s\big)\raisebox{1pt}{$^2$}}
\big({\rm Im}\,\textsf c_\phi^{\texttt P}\big)\raisebox{1pt}{$^2$} \Bigg] \,,
\end{align}
where
\begin{align}
\beta_{\hat\varsigma}^{} & \,=\, \sqrt{1-\frac{4m_\pi^2}{\hat\varsigma}} \,, &
\hat\varsigma & \,=\, (p_0^{}+p_-^{})^2 \,=\, (p_1^{}+p_2^{})^2 \,, &
\tilde\lambda_K^{} & \,=\, {\cal K}\big(m_K^2,\hat s,\hat\varsigma\big) \,,
\end{align}
and $m_\pi^{}$ in the $K^-$ $(K_L)$ formula is the isospin-average (neutral) pion mass.
Their integration ranges are \,$4m_\phi^2\le\hat s\le(m_{K^-,K^0}-2m_\pi)^2$\, and
\,$4m_\pi^2\le\hat\varsigma\le\big(m_{K^-,K^0}^{}-\hat s{}^{1/2}\big)\raisebox{1pt}{$^2$}$,\,
respectively.

Evidently, \,$K_{L,S}\to\phi\bar\phi$\, are sensitive to $\textsf c_\phi^{\texttt P}$ alone,
while \,$K\to\pi\pi'\phi\bar\phi$\, and \,$\Omega^-\to\Xi^-\phi\bar\phi$\, can probe only
the two parity-odd coefficients, $\textsf c_\phi^{\texttt A}$ and $\textsf c_\phi^{\texttt P}$,
in our approximation of the hadronic matrix elements.
On the other hand, \,$K\to\pi\phi\bar\phi$\, are sensitive to both of the parity-even
coefficients, $\textsf c_\phi^{\texttt V}$~and $\textsf c_\phi^{\texttt S}$,
but not to $\textsf c_\phi^{\texttt A,\texttt P}$.
In contrast, the spin-1/2 hyperon modes \,$\mathfrak B\to\mathfrak B'\phi\bar\phi$\, can probe
all the coefficients, according to Eq.\,(\ref{Gamma'}).
It follows that quests for \,$\mathfrak B\to\mathfrak B'\slashed E$\, as well as
\,$\Omega^-\to\Xi^-\slashed E$\, are beneficial because the acquired data could reveal
information on \,$ds\phi\phi$\, interactions which complements that gained from kaon measurements.
Furthermore, a couple of the hyperon modes, namely \,$\Sigma^+\to p\phi\bar\phi$\, and
\,$\Omega^-\to\Xi^-\phi\bar\phi$,\, can cover broader $m_\phi$ ranges than
\,$K\to\pi\pi'\phi\bar\phi$\, and therefore provide the only window into $\textsf c_\phi^{\texttt A}$
for certain $m_\phi$ values, as we discuss later on.

\section{Numerical results\label{numeric}}

Although there is still no empirical information on the hyperon decays with missing energy,
their first data may become available from BESIII not long from now.
Their branching fractions in the SM have been estimated to be less than $10^{-11}$ in
Ref.\,\cite{Tandean:2019tkm} and hence are several orders of magnitude below the estimated
sensitivity levels of BESIII \cite{Li:2016tlt}, as can be viewed in Table \ref{smB2B'nunu}.
Although the SM numbers are unlikely to be probed in the near future, NP effects turn out to be
presently allowed to boost the branching fractions tremendously to values
which may be testable by BESIII.

\begin{table}[b] \medskip
\begin{tabular}{|c||c|c|c|c|c|c|} \hline
Decay mode & \,$\Lambda\to n\nu\bar\nu$\, & \,$\Sigma^+\to p\nu\bar\nu$\, &
\,$\Xi^0\to\Lambda\nu\bar\nu$\, & \,$\Xi^0\to\Sigma^0\nu\bar\nu$\, &
\,$\Xi^-\to\Sigma^-\nu\bar\nu$\, & \,$\Omega^-\to\Xi^-\nu\bar\nu\vphantom{\int^|}$\,
\\ \hline \hline
\footnotesize $\begin{array}{c}\rm SM ~branching\vspace{-1ex}\\ \rm
fraction~\mbox{\cite{Tandean:2019tkm}}\end{array}$ &
\,$7.1\times10^{-13}$\, & \,$4.3\times10^{-13}$\, & \,$6.3\times10^{-13}$\, &
$1.0\times10^{-13}$ & $1.3\times10^{-13}$ & $4.9\times10^{-12}$
\\ \hline
\footnotesize $\begin{array}{c}\rm Expected~BESIII\vspace{-1ex}\\
\rm sensitivity~\mbox{\cite{Li:2016tlt}}\end{array}$ &
$3\times10^{-7}$ & $4\times10^{-7}$ & $8\times10^{-7}$\, & $9\times10^{-7}$ & {\bf---} &
$2.6\times10^{-5}$ \\ \hline
\end{tabular}
\caption{The branching fractions of \,$|\Delta S|=1$\, hyperon decays with missing energy in
the SM \cite{Tandean:2019tkm} and the corresponding expected sensitivities
of BESIII \cite{Li:2016tlt}.\label{smB2B'nunu}}
\end{table}

Among their kaon counterparts, the data on \,$K\to\pi\slashed E$\, are the closest to their
SM expectations.
Particularly, the E949 \cite{Artamonov:2008qb} and KOTO \cite{Ahn:2018mvc} findings,
\,${\mathcal B}(K^+\to\pi^+\nu\nu)=1.7(1.1)\times10^{-10}$ \cite{Tanabashi:2018oca}
and \,${\mathcal B}\big(K_L\to\pi^0\nu\bar\nu\big) < 3.0\times10^{-9}$ at 90\% confidence level
(CL), respectively, are not far from the SM numbers
\,${\mathcal B}(K^+\to\pi^+\nu\nu)=\big(8.5_{-1.2}^{+1.0}\big)\times10^{-11}$ and
\,${\mathcal B}(K_L\to\pi^0\nu\bar\nu)=\big(3.2_{-0.7}^{+1.1}\big)\times10^{-11}$~\cite{Bobeth:2017ecx}.
We can regard the upper ends of the 90\%-CL ranges of their deviations from the SM predictions,
\begin{align} \label{K2pinv}
\Delta{\mathcal B}(K^-\to\pi^-\slashed E) & \,<\, 2.7\times10^{-10} \,, &
\Delta{\mathcal B}(K_L\to\pi^0\slashed E) & \,<\, 3.0\times10^{-9}  \,, &
\end{align}
as capping the NP contributions.
These conditions prevent $\textsf c_\phi^{\texttt V,\texttt S}$ from being sizable.

By contrast, there is still comparatively greater room for NP to impact
$K\to\slashed E$ and $K\to\pi\pi'\slashed E$.
For the former, although direct searches are yet to be conducted \cite{Tanabashi:2018oca}, there
are indirect upper bounds (at 95\% CL) extracted from the existing data on $K_{L,S}$ visible decay
channels \cite{Gninenko:2014sxa}:
\begin{align} \label{K2inv}
{\mathcal B}(K_L\to\slashed E) & \,<\, 6.3\times10^{-4} \,, &
{\mathcal B}(K_S\to\slashed E) & \,<\, 1.1\times10^{-4} \,. &
\end{align}
These numbers are way higher than
\,${\cal B}(K_L\to\nu\bar\nu)\raisebox{1pt}{\footnotesize\,$\lesssim$\,}1\times10^{-10}$\, and
\,${\cal B}(K_S\to\nu\bar\nu)\raisebox{1pt}{\footnotesize\,$\lesssim$\,}2\times10^{-14}$
estimated in the SM supplemented with neutrino
mass \cite{Tandean:2019tkm,Gninenko:2014sxa}.\footnote{These results follow from
the use of the neutrino mass' highest direct limit \,$m_{\nu_\tau}^{\rm exp}<18.2$\,MeV\, listed
by the Particle Data Group \cite{Tanabashi:2018oca}.
If one employs instead the cosmological bound of under 1 eV for the sum of neutrino masses
\cite{Tanabashi:2018oca}, the invisible widths of $K_{L,S}$ turn out to be dominated by their
decays into four neutrinos with branching fractions below $10^{-21}$ and $10^{-24}$,
respectively \cite{Bhattacharya:2018msv}.}
For the four-body modes, the empirical limits
\,${\mathcal B}(K^-\to\pi^0\pi^-\nu\bar\nu)<4.3\times10^{-5}$~\cite{Adler:2000ic} and
\,${\mathcal B}(K_L\to\pi^0\pi^0\nu\bar\nu)<8.1\times10^{-7}$~\cite{E391a:2011aa} are the only
data available and again vastly exceed the SM expectations of order $10^{-14}$ and $10^{-13}$,
respectively \cite{Kamenik:2011vy,Littenberg:1995zy,Chiang:2000bg}.
Accordingly, we may impose
\begin{align} \label{K2ppinv}
{\mathcal B}(K^-\to\pi^0\pi^-\slashed E) & \,<\, 4\times10^{-5} \,, &
{\mathcal B}(K_L\to\pi^0\pi^0\slashed E) & \,<\, 8\times10^{-7}  &
\end{align}
on NP contributions.

To explore how much NP can enhance the hyperon rates, we consider first the possibility
that the $\phi$ mass can be ignored relative to the pion mass or vanishes, \,$m_\phi=0$.\,
In this case, integrating the differential rates of the baryon decays,
we arrive at the branching fractions
\begin{align} \label{B2B'ff}
{\cal B}\big(\Lambda\to n\phi\bar\phi\big) & \,=\, \Big[
0.11\, \big|\textsf c_\phi^{\texttt V}\big|^2 + 0.18\, \big|\textsf c_\phi^{\texttt A}\big|^2
+ \Big( 42\, \big|\textsf c_\phi^{\texttt S}\big|^2
+ 12\, \big|\textsf c_\phi^{\texttt P}\big|^2 \Big) {\rm GeV}^{-2} \Big] 10^8 \rm~GeV^4 \,,
\nonumber \\
{\cal B}\big(\Sigma^+\to p\phi\bar\phi\big) & \,=\, \Big[
0.13\, \big|\textsf c_\phi^{\texttt V}\big|^2 + 0.046\, \big|\textsf c_\phi^{\texttt A}\big|^2
+ \Big( 49\, \big|\textsf c_\phi^{\texttt S}\big|^2
+ 3.4\, \big|\textsf c_\phi^{\texttt P}\big|^2 \Big) {\rm GeV}^{-2} \Big] 10^8 \rm~GeV^4 \,,
\nonumber \\
{\cal B}\big(\Xi^0\to\Lambda\phi\bar\phi\big) & \,=\, \Big[
0.23\, \big|\textsf c_\phi^{\texttt V}\big|^2 + 0.025\, \big|\textsf c_\phi^{\texttt A}\big|^2
+ \Big(91\, \big|\textsf c_\phi^{\texttt S}\big|^2
+ 1.8\, \big|\textsf c_\phi^{\texttt P}\big|^2 \Big) {\rm GeV}^{-2} \Big] 10^8 \rm~GeV^4 \,,
\end{align}
\begin{align}
{\cal B}\big(\Xi^0\to\Sigma^0\phi\bar\phi\big) & \,=\, \Big[
0.07\, \big|\textsf c_\phi^{\texttt V}\big|^2 + 0.34\, \big|\textsf c_\phi^{\texttt A}\big|^2
+ \Big( 28\, \big|\textsf c_\phi^{\texttt S}\big|^2
+ 23\, \big|\textsf c_\phi^{\texttt P}\big|^2 \Big) {\rm GeV}^{-2} \Big] 10^7 \rm~GeV^4 \,,
\nonumber \\ \label{X2Sff}
{\cal B}\big(\Xi^-\to\Sigma^-\phi\bar\phi\big) & \,=\, \Big[
0.09\, \big|\textsf c_\phi^{\texttt V}\big|^2 + 0.42\, \big|\textsf c_\phi^{\texttt A}\big|^2
+ \Big( 33\, \big|\textsf c_\phi^{\texttt S}\big|^2
+ 28\, \big|\textsf c_\phi^{\texttt P}\big|^2 \Big) {\rm GeV}^{-2} \Big] 10^7 \rm~GeV^4 \,,
\end{align}
\begin{align} \label{BO2Xff}
{\cal B}\big(\Omega^-\to\Xi^-\phi\bar\phi\big) & \,=\, \Big(
2.0\, \big|\textsf c_\phi^{\texttt A}\big|^2
+ 152\, \big|\textsf c_\phi^{\texttt P}\big|^2\, {\rm GeV}^{-2} \Big) 10^8 \rm~GeV^4 \,. &
\end{align}
All these results have incorporated the form factors mentioned in Sec.\,\ref{hyperons}.

As mentioned above, the current \,$K\to\pi\nu\bar\nu$\, data
do not leave ample room for NP to affect $\textsf c_\phi^{\texttt V,\texttt S}$ significantly.
More specifically, if these coefficients are the only ones being nonzero, comparing
their contributions, for \,$m_\phi=0$,\,
\begin{align} \label{BK2piff}
{\cal B}\big(K^-\to\pi^-\phi\bar\phi\big) & \,=\, \Big( 0.033\,
\big|\textsf c_\phi^{\texttt V}\big|^2 + 14\, \big|\textsf c_\phi^{\texttt S}\big|^2
{\rm\,GeV}^{-2} \Big) 10^{11} \rm\,GeV^4 \,,
\nonumber \\
{\cal B}\big(K_L^{}\to\pi^0\phi\bar\phi\big) & \,=\, \Big[ 0.15\, \big({\rm Im}\,
\textsf c_\phi^{\texttt V}\big)^2 + 59\, \big({\rm Re}\, \textsf c_\phi^{\texttt S}\big)^2
{\rm\,GeV}^{-2} \Big] 10^{11} \rm\,GeV^4 &
\end{align}
with the corresponding bounds in Eq.\,(\ref{K2pinv}), we find that the resulting hyperon branching
fractions in Eqs.\,\,(\ref{B2B'ff}) and (\ref{X2Sff}) are below $10^{-11}$.
This scenario would therefore be out of BESIII reach according to
Table \ref{smB2B'nunu}.

Therefore, hereafter we concentrate on the possibility that NP can generate considerable effects
solely through $\textsf c_\phi^{\texttt A,\texttt P}$.
With this assumption and \,$m_\phi=0$,\, for the two-body kaon decays we get
\begin{align} \label{BK2ff}
{\cal B}\big(K_L^{}\to\phi\bar\phi\big) & =\, 5.9\!\times\!10^{14} \rm\, GeV^2\,
\big({\rm Im}\, \textsf c_\phi^{\texttt P}\big)^2 \,, &
\nonumber \\
{\cal B}\big(K_S^{}\to\phi\bar\phi\big) & =\, 1.0\!\times\!10^{12} \rm\, GeV^2\,
\big({\rm Re}\, \textsf c_\phi^{\texttt P}\big)^2 \,,
\end{align}
implying
\begin{align} \label{tcstcp}
\big|\textsf c_\phi^{\texttt P}\big|\raisebox{2pt}{$^2$} & \,=\,
\frac{1.7\times10^{-15}}{\rm GeV^2}\, {\cal B}\big(K_L^{}\to\phi\bar\phi\big)
+ \frac{9.7\times10^{-13}}{\rm GeV^2}\, {\cal B}\big(K_S^{}\to\phi\bar\phi\big) \,, &
\end{align}
and for the four-body decays
\begin{align} \label{BK2pipiff}
{\cal B}\big(K^-\to\pi^-\pi^0\phi\bar\phi\big) & \,=\, \Big( 0.16\,
\big|\textsf c_\phi^{\texttt A}\big|^2 + 14\, \big|\textsf c_\phi^{\texttt P}\big|^2
{\rm\,GeV}^{-2} \Big) 10^6 \rm\,GeV^4 \,,
\nonumber \\
{\cal B}\big(K_L^{}\to\pi^0\pi^0\phi\bar\phi\big) & \,=\, \Big[ 0.21\, \big({\rm Re}\,
\textsf c_\phi^{\texttt A}\big)^2 + 64\, \big({\rm Im}\, \textsf c_\phi^{\texttt P}\big)^2
{\rm\,GeV}^{-2} \Big] 10^7 \rm\;GeV^4 \,. &
\end{align}
To illustrate the potential impact of sizable $\textsf c_\phi^{\texttt A,\texttt P}$ on
the hyperon modes, we can look at a couple of examples with different choices of
the nonvanishing coupling, which we take to be real to avoid any new $CP$-violation source.

Thus, if only \,$\textsf c_\phi^{\texttt P}\neq0$,\, the \,$K\to\slashed E$\, restraints in
Eq.\,(\ref{K2inv}) turn out to be more stringent than the \,$K\to\pi\pi'\slashed E$\, ones
in Eq.\,(\ref{K2ppinv}) and, with the aid of Eq.\,(\ref{tcstcp}), lead to
\,$\big|\textsf c_\phi^{\texttt P}\big|\raisebox{1pt}{$^2$}<1.1\times10^{-16}{\rm\;GeV}^{-2}$.\,
Plugging this into Eqs.\,\,(\ref{B2B'ff})-(\ref{BO2Xff}) and setting
\,$\textsf c_\phi^{\texttt V,\texttt A,\texttt S}=0$,\, we obtain
\begin{align}
{\cal B}\big(\Lambda\to n\phi\bar\phi\big)  & \,<\, 1.3\times10^{-7} \,, &
{\cal B}\big(\Sigma^+\to p\phi\bar\phi\big) & \,<\, 3.7\times10^{-8} \,, &
\nonumber \\
{\cal B}\big(\Xi^0\to\Lambda\phi\bar\phi\big)  & \,<\, 1.9\times10^{-8} \,, &
{\cal B}\big(\Xi^0\to\Sigma^0\phi\bar\phi\big) & \,<\, 2.5\times10^{-8} \,, &
\nonumber \\
{\cal B}\big(\Xi^-\to\Sigma^-\phi\bar\phi\big) & \,<\, 3.0\times10^{-8} \,, &
{\cal B}\big(\Omega^-\to\Xi^-\phi\bar\phi\big) & \,<\, 1.6\times10^{-6} \,. &
\end{align}
The upper ends of these ranges greatly exceed the corresponding SM values quoted in
Table\,\,\ref{smB2B'nunu} but are still below the expected BESIII sensitivity levels,
mostly by at least an order of magnitude.

If instead $\textsf c_\phi^{\texttt A}$ is nonvanishing and real, whereas
\,$\textsf c_\phi^{\texttt V,\texttt P,\texttt S}=0$,\, then the \,$K\to\slashed E$\,
bounds do not matter any more but the \,$K\to\pi\pi'\slashed E$\, ones in Eq.\,(\ref{K2ppinv})
still do, the \,$K_L\to\pi^0\pi^0\slashed E$\, one being the stronger and yielding
\,$\big({\rm Re}\,\textsf c_\phi^{\texttt A}\big)^2<3.8\times10^{-13}{\rm\;GeV}^{-4}$.\,
This now translates into
\begin{align} \label{hypBRmax}
{\cal B}\big(\Lambda\to n\phi\bar\phi\big) & \,<\, 6.6\times10^{-6} \,, &
{\cal B}\big(\Sigma^+\to p\phi\bar\phi\big) & \,<\, 1.7\times10^{-6} \,, &
\nonumber \\
{\cal B}\big(\Xi^0\to\Lambda\phi\bar\phi\big) & \,<\, 9.4\times10^{-7} \,, &
{\cal B}\big(\Xi^0\to\Sigma^0\phi\bar\phi\big) & \,<\, 1.3\times10^{-6} \,, &
\nonumber \\
{\cal B}\big(\Xi^-\to\Sigma^-\phi\bar\phi\big) & \,<\, 1.6\times10^{-6} \,, &
{\cal B}\big(\Omega^-\to\Xi^-\phi\bar\phi\big) & \,<\, 7.5\times10^{-5} \,. &
\end{align}
Their upper values are larger than the corresponding BESIII sensitivity
levels quoted in Table\,\,\ref{smB2B'nunu}.
Allowing \,${\rm Im}\,\textsf c_\phi^{\texttt A}\neq0$,\, one could get higher branching
fractions with $\textsf c_\phi^{\texttt A}$ being complex, as its imaginary part would escape
the \,$K_L\to\pi^0\pi^0\slashed E$\, restriction and be subject only to the milder
\,$K^-\to\pi^0\pi^-\slashed E$\, one.
The numbers in Eq.\,(\ref{hypBRmax}) are the same as their counterparts in the case treated in
Ref.\,\cite{Tandean:2019tkm} where the invisible particles are Dirac spin-1/2 fermions,
highlighting the similarities of the two scenarios when the masses of the invisible particles
can be neglected.

\begin{figure}[b!] \bigskip
\includegraphics[width=9cm]{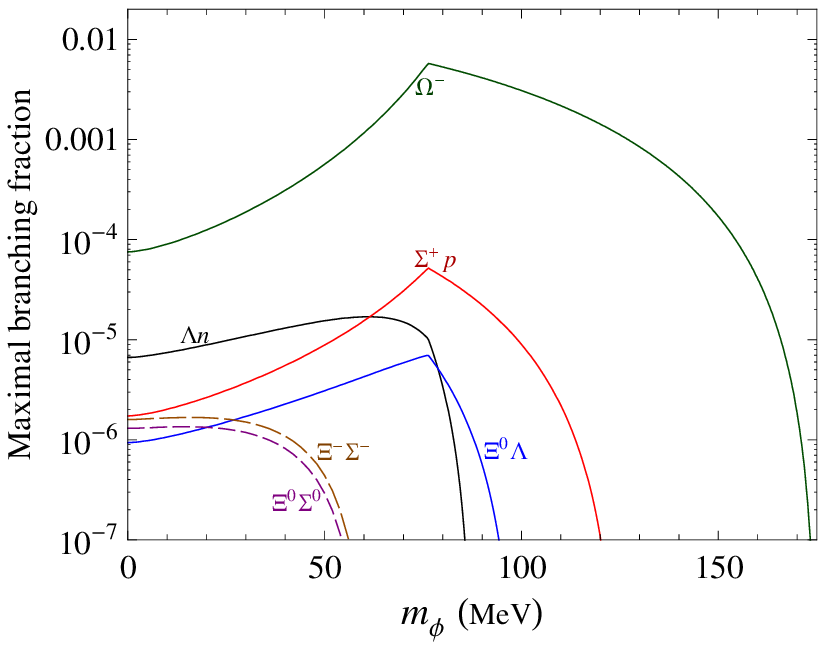}\vspace{-3pt}
\caption{The maximal branching fractions of \,$\mathfrak B\to\mathfrak B'\phi\bar\phi$\, with
\,$\mathfrak{BB}'=\Lambda n,\Sigma^+p,\Xi^0\Lambda,\Xi^0\Sigma^0,\Xi^-\Sigma^-$\, and
of~\,$\Omega^-\to\Xi^-\phi\bar\phi$,\, indicated on the plot by the $\mathfrak{BB}'$
and $\Omega^-$\, labels, respectively, versus $m_\phi$, induced by the contribution of
${\rm Re}\,\textsf c_\phi^{\texttt A}$ alone, subject to the \,$K_L\to\pi^0\pi^0\slashed E$\,
constraint and the perturbativity requirement for
\,$m_\phi>76$\,\,MeV\, as explained in the text.}\label{Bphi}
\end{figure}

We consider next the possibility that $m_\phi$ is nonnegligible and let it vary over
the allowed kinematical range of each of the hyperon decays.
As in the second example, we suppose that $\textsf c_\phi^{\texttt A}$ is nonzero and real,
while \,$\textsf c_\phi^{\texttt V,\texttt S,\texttt P}=0$,\, and so the
\,$K_L\to\pi^0\pi^0\slashed E$\, restriction again needs to be taken into account.
Now, as $m_\phi$ grows this restriction increasingly weakens until it no longer applies at
\,$m_\phi\simeq114$\,\,MeV\, when \,$K_L\to\pi^0\pi^0\slashed E$\, becomes kinematically closed.
Accordingly, for \,$m_\phi>0$\, the hyperon rates can be substantially bigger than their
\,$m_\phi=0$\, values, and for \,$m_\phi>114$\,\,MeV\, the kaon decay constraint is no more.
However, as the restraint on $\big|{\rm Re}\,\textsf c_\phi^{\texttt A}\big|$ continues to loosen,
its rising upper bound may no longer be compatible with the perturbativity of the $\phi$
interactions.
To ensure that they remain perturbative, inspired by the form of the standard effective
weak interaction we impose
\,$\big|{\rm Re}\,\textsf c_\phi^{\texttt A}\big|<G_{\rm F}/\sqrt2\simeq\rm8.2/TeV^2$,\,
where $G_{\rm F}$ is the Fermi constant.
Since this corresponds to the requisite
\,\mbox{${\cal B}\big(K_L^{}\to\pi^0\pi^0\phi\bar\phi\big)<8\times10^{-7}$}\, at
\,$m_\phi\simeq76$\,\,MeV,\, we then set
\,$\big|{\rm Re}\,\textsf c_\phi^{\texttt A}\big|=G_{\rm F}/\sqrt2$\, for~\,$m_\phi>76$\,\,MeV.\,
In Fig.\,\ref{Bphi} we depict the resulting maximal branching fractions of the hyperon modes versus
$m_\phi$ varying from 0 to 175\,MeV, which is approximately its highest value in
\,$\Omega^-\to\Xi^-\phi\bar\phi$.\,
Evidently, over the majority of their respective $m_\phi$ ranges, four of the modes (the ones with
\,$\mathfrak{BB}'=\Lambda n,\Sigma^+p,\Xi^0\Lambda$\, and the $\Omega^-$ one) have maximal branching
fractions that are significantly greater than their \,$m_\phi=0$\, values, implying that
their \,$m_\phi>0$\, predictions are easier to test experimentally.

\begin{table}[b!] \bigskip
\begin{tabular}{|c||c|c||c|c|} \hline
~Kaon mode~ & ~$K\to\phi\bar\phi$~ & ~$K\to\pi\pi'\phi\bar\phi$~ &
\,$K\to\texttt{\textsl f}\bar{\texttt{\textsl f}}$\, &
\,$K\to\pi\pi'\texttt{\textsl f}\bar{\texttt{\textsl f}}\vphantom{\int^|}$\,
\\ \hline \hline
Couplings & $\textsf c_\phi^{\texttt P}\vphantom{\int_|^|}$ &
$\textsf c_\phi^{\texttt A}$, $\textsf c_\phi^{\texttt P}$ &
~$\tilde{\textsf c}_{\texttt{\textsl f}}^{\textsc a}$,
$\tilde{\textsf c}_{\texttt{\textsl f}}^{\textsc s}$,
$\tilde{\textsf c}_{\texttt{\textsl f}}^{\textsc p}$~  &
~$\tilde{\textsf c}_{\texttt{\textsl f}}^{\textsc v}$,
$\tilde{\textsf c}_{\texttt{\textsl f}}^{\textsc a}$,
$\tilde{\textsf c}_{\texttt{\textsl f}}^{\textsc s}$,
$\tilde{\textsf c}_{\texttt{\textsl f}}^{\textsc p}$~
\\ \hline
\end{tabular}
\caption{New-physics couplings contributing to \,$K\to\slashed E$\, and
\,$K\to\pi\pi'\slashed E$\, if $\slashed E$ is carried away by spin-0 bosons $\phi\bar\phi$
or Dirac spin-1/2 fermions $\texttt{\textsl f}\bar{\texttt{\textsl f}}$ and their
masses are nonzero, \,$m_{\phi,\texttt{\textsl f}}>0$.\,
All these couplings belong to operators involving parity-odd $ds$ quark
bilinears.\label{phi-vs-f}}
\end{table}

\begin{figure}[b] \bigskip
\includegraphics[width=9cm]{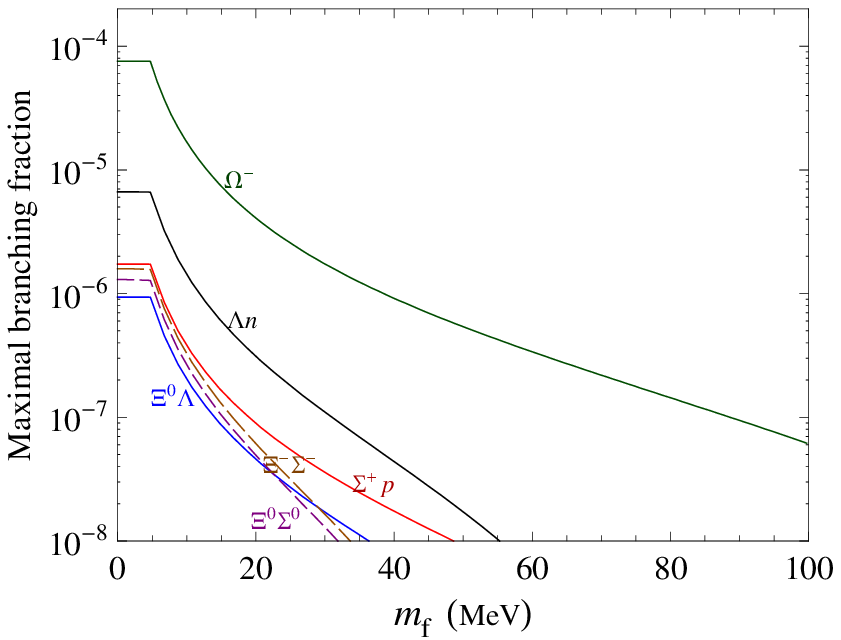}\vspace{-3pt}
\caption{The maximal branching fractions of
\,$\mathfrak B\to\mathfrak B'\texttt{\textsl f}\bar{\texttt{\textsl f}}$\, with
\,$\mathfrak{BB}'=\Lambda n,\Sigma^+p,\Xi^0\Lambda,\Xi^0\Sigma^0,\Xi^-\Sigma^-$\, and
of~\,$\Omega^-\to\Xi^-\texttt{\textsl f}\bar{\texttt{\textsl f}}$\,
versus $m_{\texttt{\textsl f}}^{}$, induced by the contribution of
${\rm Re}\,\tilde{\textsf c}_{\texttt{\textsl f}}^{\textsc a}$ alone, subject to
the \,$K_L\to\pi^0\pi^0\slashed E$\, and \,$K_L\to\slashed E$\, constraints, with the latter
becoming more important for \,$m_{\texttt{\textsl f}}^{}>5$\,\,MeV.}\label{Bf}
\end{figure}

Now, in the \,$s\to d\slashed E$\, scenario explored in Ref.\,\cite{Tandean:2019tkm} the invisible
particles are Dirac spin-1/2 fermions, $\texttt{\textsl f}\bar{\texttt{\textsl f}}$, and
the numerical work therein focused on instances where the $\texttt{\textsl f}$ mass, $m_{\textsl{\texttt f}}$, was negligible or zero.
Here we look at what may happen if $m_{\textsl{\texttt f}}$ is not negligible.
In that case, as Table\,\,\ref{phi-vs-f} summarizes,
\,$K\to\texttt{\textsl f}\bar{\texttt{\textsl f}}$\, and
\,$K\to\pi\pi'\texttt{\textsl f}\bar{\texttt{\textsl f}}$\, receive contributions from the NP
couplings $\tilde{\textsf c}_{\texttt{\textsl f}}^{\textsc a,\textsc s,\textsc p}$ and
$\tilde{\textsf c}_{\texttt{\textsl f}}^{\textsc v,\textsc a,\textsc s,\textsc p}$, respectively,
which parametrize the effective $ds\texttt{\textsl f}_{\!}\texttt{\textsl f}$ interactions
described by the Lagrangian
\,${\cal L}_{\texttt{\textsl f}}\supset -\overline{d} \gamma^\eta\gamma_5^{} s\, \overline{\texttt{\textsl f}} \gamma_\eta^{}
\big( \tilde{\textsf c}_{\texttt{\textsl f}}^{\textsc v}
+ \gamma_5^{} \tilde{\textsf c}{}_{\texttt{\textsl f}}^{\textsc a} \big) \texttt{\textsl f}
- \overline{d} \gamma_5^{}s\, \overline{\texttt{\textsl f}}
\big( \tilde{\textsf c}_{\texttt{\textsl f}}^{\textsc s} + \gamma_5^{}
\tilde{\textsf c}_{\texttt{\textsl f}}^{\textsc p} \big) \texttt{\textsl f} + {\rm H.c.}$\,
used in Ref.\,\cite{Tandean:2019tkm}.\footnote{In Ref.\,\cite{Tandean:2019tkm}, an overall factor
of \,$\beta=\sqrt{1-4m_{\textsl{\texttt f}}^2/\hat s}$\, is missing from both Eqs.\,(B.3) and (B.4)
for the double-differential rates of \,$K^-\to\pi^0\pi^-\texttt{\textsl f}\bar{\texttt{\textsl f}}$\,
and \,$K_L\to\pi^0\pi^0\texttt{\textsl f}\bar{\texttt{\textsl f}}$.\,
It appears to be missing also from the corresponding formulas in~Ref.\,\cite{Kamenik:2011vy}.}
Consequently, for $\tilde{\textsf c}_{\texttt{\textsl f}}^{\textsc a,\textsc s,\textsc p}$
the \,$K\to\slashed E$\, restrictions become more important than
the \,$K\to\pi\pi'\slashed E$\, ones over most of the \,$m_{\textsl{\texttt f}}^{}>0$\,
region, with the implication that the resulting hyperon rates tend to be much smaller than
their \,$m_{\textsl{\texttt f}}^{}=0$\, values.
This is illustrated in Fig.\,\ref{Bf}, for which
${\rm Re}\,\tilde{\textsf c}_{\texttt{\textsl f}}^{\textsc a}$ is taken to be the only coupling
being present.
From the figure, we learn that the \,$K_L\to\slashed E$\, constraint is stricter than
the \,$K_L\to\pi^0\pi^0\slashed E$\, one if \,$m_{\texttt{\textsl f}}^{}>5$\,\,MeV.\,
This plot also reveals striking differences from Fig.\,\ref{Bphi} when the invisible particle's
mass is not negligible.

\begin{figure}[t!]
\includegraphics[width=9cm]{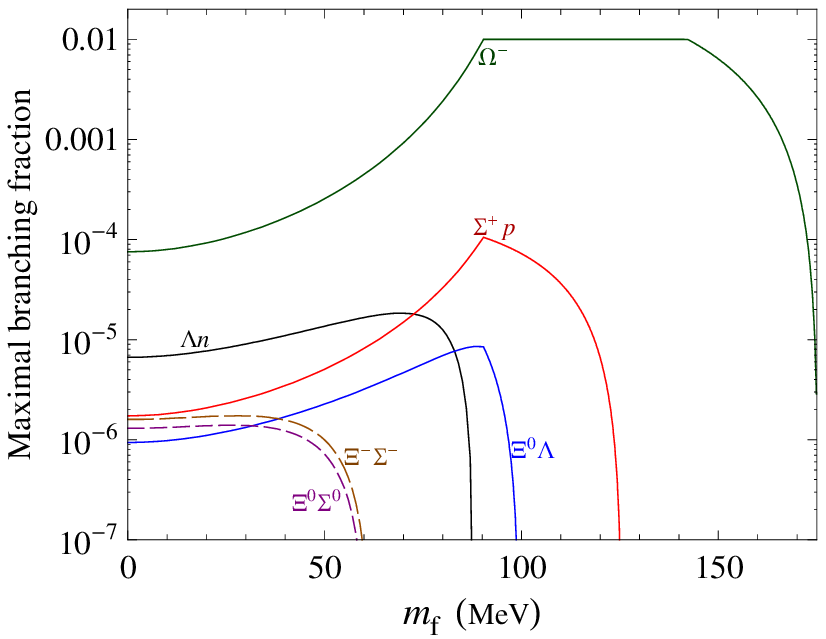}\vspace{-3pt}
\caption{The maximal branching fractions of
\,$\mathfrak B\to\mathfrak B'\texttt{\textsl f}\bar{\texttt{\textsl f}}$\, with
\,$\mathfrak{BB}'=\Lambda n,\Sigma^+p,\Xi^0\Lambda,\Xi^0\Sigma^0,\Xi^-\Sigma^-$\, and
of~\,$\Omega^-\to\Xi^-\texttt{\textsl f}\bar{\texttt{\textsl f}}$\,
versus $m_{\texttt{\textsl f}}^{}$, induced by the contribution of
${\rm Re}\,\tilde{\textsf c}_{\texttt{\textsl f}}^{\textsc v}$ alone,
subject to the \,$K_L\to\pi^0\pi^0\slashed E$\, constraint and the perturbativity and
$\Omega^-$ data requirements for \,$m_{\texttt{\textsl f}}^{}>90$\,\,MeV\, as explained in
the text.}\label{Bf-tcv}
\end{figure}

If instead only ${\rm Re}\,\tilde{\textsf c}_{\texttt{\textsl f}}^{\textsc v}$ is nonvanishing,
then it evades the \,$K\to\slashed E$\, restrictions completely and is subject only to
the \,$K\to\pi\pi'\slashed E$\, ones, similarly to the bosonic scenario with
$\textsf c_\phi^{\texttt A}$ being the only coupling,
as can be viewed also from Table\,\,\ref{phi-vs-f}.
We again demand
\,$|{\rm Re}\,\tilde{\textsf c}_{\texttt{\textsl f}}^{\textsc v}|<G_{\rm F}/\sqrt2$\, to
satisfy the perturbativity requisite, which in this case applies for
\,$m_{\textsl{\texttt f}}^{}>96$\,\,MeV.\,
However, under these conditions we find that
${\rm Re}\,\tilde{\textsf c}_{\texttt{\textsl f}}^{\textsc v}$ for
\,$m_{\textsl{\texttt f}}^{}\in[90,142]$\,\,MeV\, translates into upper bounds on
${\cal B}\big(\Omega^-\to\Xi^-\texttt{\textsl f}\bar{\texttt{\textsl f}}\big)$ that exceed
the allowed value of roughly 1\% for the unknown $\Omega^-$ decay channels,
which is inferred from the data on the observed ones \cite{Tanabashi:2018oca}.
Therefore, for this $m_{\textsl{\texttt f}}^{}$ region we additionally require
${\rm Re}\,\tilde{\textsf c}_{\texttt{\textsl f}}^{\textsc v}$ to yield
\,${\cal B}\big(\Omega^-\to\Xi^-\texttt{\textsl f}\bar{\texttt{\textsl f}}\big)<1$\%.\,
We display the resulting branching-fraction limits of the various hyperon modes in
Fig.\,\ref{Bf-tcv} which shows stark differences from Fig.\,\ref{Bf} for
\,$m_{\texttt{\textsl f}}^{}>5$\,\,MeV.\,
On the other hand, Fig.\,\ref{Bf-tcv} resembles Fig.\,\ref{Bphi} in important ways,
especially in their depictions of the maximal branching fractions which are hugely amplified
over much of the \,$m_{\phi,\texttt{\textsl f}}>0$\, ranges relative to their values
at \,$m_{\phi,\texttt{\textsl f}}=0$.\,
There are also dissimilarities between Figs.\,\,\ref{Bphi} and \ref{Bf-tcv}, one of which is
that the curves in the latter tend to be higher at larger masses.

Finally, we note that for $\texttt{\textsl f}$ being a Majorana, rather than Dirac, fermion
the $ds\texttt{\textsl f}_{\!}\texttt{\textsl f}$ operators involving the vector bilinear of
$\texttt{\textsl f}$ do not exist, as its Majorana nature implies
\,$\overline{\texttt{\textsl f}}\gamma^\mu\texttt{\textsl f}=0$ \cite{Dreiner:2008tw}.
In other words, the contribution of $\tilde{\textsf c}_{\texttt{\textsl f}}^{\textsc v}$
(and of $\texttt C_{\texttt{\textsl f}}^{\texttt V}$, which belongs to
the \,$\overline d\gamma^\eta s\, \overline{\texttt{\textsl f}}\gamma_\eta^{}\texttt{\textsl f}$\,
operator \cite{Tandean:2019tkm}) is absent if $\texttt{\textsl f}$ is
a Majorana particle.
It follows that in this case the substantial increases of the maximal branching fractions like
those seen in Fig.\,\ref{Bf-tcv} do not occur and instead, if significant NP affects mostly
${\rm Re}\,\tilde{\textsf c}_{\texttt{\textsl f}}^{\textsc a}$, the situation is roughly
similar to that pictured in Fig.\,\ref{Bf}.

\section{Conclusions\label{concl}}

We have explored the strangeness-changing decays of hyperons into another baryon plus missing energy
carried away by a pair of invisible spinless bosons, $\phi\bar\phi$, which are SM gauge singlet.
These processes arise from their effective interactions with the $d$ and $s$ quarks at low energies.
Adopting a model-independent approach, we start from an effective Lagrangian respecting SM gauge
invariance and containing dimension-six $ds\phi\phi$ operators.
Thus, they contribute also to FCNC kaon decays with missing energy and are subject
to restrictions from their data.
Although the existing \,$K\to\pi\slashed E$\, constraints do not permit NP to generate sizable
effects via parity-even $ds\phi\phi$ operators, we demonstrate that the restrictions from
\,$K\to\slashed E$\, and \,$K\to\pi\pi'\slashed E$\, are much weaker on the parity-odd operators.
Specifically, the one with the axial-vector quark bilinear needs to fulfill only
the mild \,$K\to\pi\pi'\slashed E$\, restraints and could give rise to considerable hyperon
rates which may be within the reach of BESIII or future measurements.
Moreover, if NP enters predominantly through the coefficient $\textsf c_\phi^{\texttt A}$
of this operator and \,$m_\phi>0$, the kaon decay constraints become weakened and as a result
the hyperon rates may be greatly enhanced with respect to their \,$m_\phi=0$\, predictions.
Interestingly, these kaon constraints on $\textsf c_\phi^{\texttt A}$ are absent if
\,$m_\phi>114$\,\,MeV\, but two of the hyperon modes, especially \,$\Omega^-\to\Xi^-\phi\bar\phi$,\,
can still serve as direct probes of $\textsf c_\phi^{\texttt A}$ until $m_\phi$
reaches \,175\,\,MeV.\,
We also perform comparisons with the scenario in which the invisible particles are Dirac spin-1/2
fermions.
We discuss special instances in which the bosonic and fermionic cases may be discriminated
experimentally, particularly if their masses are nonvanishing.
The results of our analysis illustrate well that these rare hyperon decays and their kaon
counterparts are complementary to each other in providing access to possible NP in
\,$s\to d\slashed E$\, transitions.

\acknowledgements

This research was supported in part by the MOST (Grant No. MOST 106-2112-M-002-003-MY3).

\end{document}